\documentclass[american]{llncs}
\usepackage[T1]{fontenc}
\usepackage[latin9]{inputenc}
\usepackage{xcolor}
\usepackage{pdfcolmk}
\usepackage{refstyle}
\usepackage{graphicx}
\PassOptionsToPackage{normalem}{ulem}
\usepackage{ulem}

\makeatletter

\AtBeginDocument{\providecommand\figref[1]{\ref{fig:#1}}}
\AtBeginDocument{\providecommand\subsecref[1]{\ref{subsec:#1}}}
\RS@ifundefined{subsecref}
  {\newref{subsec}{name = \RSsectxt}}
  {}
\RS@ifundefined{thmref}
  {\def\RSthmtxt{theorem~}\newref{thm}{name = \RSthmtxt}}
  {}
\RS@ifundefined{lemref}
  {\def\RSlemtxt{lemma~}\newref{lem}{name = \RSlemtxt}}
  {}

\providecolor{lyxadded}{rgb}{0,0,1}
\providecolor{lyxdeleted}{rgb}{1,0,0}

\DeclareRobustCommand{\lyxsout}[1]{\ifx\\#1\else\sout{#1}\fi}

\makeatother

\usepackage{babel}
\begin{document}
\title{{\large{}Bringing Anatomical Information into Neuronal Network Models}}
\author{S.J. van Albada$^{1,2}$, A. Morales-Gregorio$^{1,3}$, T. Dickscheid$^{4}$,
A. Goulas$^{5}$, R. Bakker$^{1,6}$, S. Bludau$^{4}$, G. Palm$^{1}$,
C.-C. Hilgetag$^{5,7}$, and M. Diesmann$^{1,8,9}$}
\institute{$^{1}$Institute of Neuroscience and Medicine (INM-6) Computational
and Systems Neuroscience, Institute for Advanced Simulation (IAS-6)
Theoretical Neuroscience, and JARA-Institut Brain Structure-Function
Relationships (INM-10), Jülich Research Centre, Jülich, Germany\\
$^{2}$Institute of Zoology, Faculty of Mathematics and Natural Sciences,
University of Cologne, Germany\\
$^{3}$RWTH Aachen University, Aachen, Germany\\
$^{4}$Institute of Neuroscience and Medicine (INM-1) Structural and
Functional Organisation of the Brain, Jülich Research Centre, Jülich,
Germany\\
$^{5}$Institute of Computational Neuroscience, University Medical
Center Eppendorf, Hamburg, Germany\\
$^{6}$Department of Neuroinformatics, Donders Centre for Neuroscience,
Radboud University, Nijmegen, the Netherlands\\
$^{7}$Department of Health Sciences, Boston University, Boston, USA\\
$^{8}$Department of Psychiatry, Psychotherapy and Psychosomatics,
School of Medicine, RWTH Aachen University, Aachen, Germany\\
$^{9}$Department of Physics, Faculty 1, RWTH Aachen University, Aachen,
Germany}
\maketitle
\begin{abstract}
For constructing neuronal network models computational neuroscientists
have access to wide-ranging anatomical data that nevertheless tend
to cover only a fraction of the parameters to be determined. Finding
and interpreting the most relevant data, estimating missing values,
and combining the data and estimates from various sources into a coherent
whole is a daunting task. With this chapter we aim to provide guidance
to modelers by describing the main types of anatomical data that may
be useful for informing neuronal network models. We further discuss
aspects of the underlying experimental techniques relevant to the
interpretation of the data, list particularly comprehensive data sets,
and describe methods for filling in the gaps in the experimental data.
Such methods of `predictive connectomics' estimate connectivity where
the data are lacking based on statistical relationships with known
quantities. It is instructive, and in certain cases necessary, to
use organizational principles that link the plethora of data within
a unifying framework where regularities of brain structure can be
exploited to inform computational models. In addition, we touch upon
the most prominent features of brain organization that are likely
to influence predicted neuronal network dynamics, with a focus on
the mammalian cerebral cortex. Given the still existing need for modelers
to navigate a complex data landscape full of holes and stumbling blocks,
it is vital that the field of neuroanatomy is moving toward increasingly
systematic data collection, representation, and publication.

\end{abstract}

\section{Introduction}

Some of the defining characteristics of a neuronal network model are
the size of the neuronal populations and the connectivity between
the neurons. To determine these properties, the modeler has access
to information in multiple forms and based on various experimental
methods, where the completeness of the data varies widely across species
and brain areas. For instance, the connectivity data for the nervous
system of the nematode (roundworm) \textit{C. Elegans} are nearly
complete and have enabled full connectomes to be derived with minimal
extrapolation from the data \cite{Cook2019_63}. These graphs encode
all connections between all of the neurons of the male and hermaphrodite
worms. However, the 302 neurons of the hermaphrodite and the 385
neurons of the male worm pale in comparison to larger brains such
as the human brain with its roughly 86 billion neurons and trillions
of connections. Here, and for most species, measuring a full connectome
is still far from feasible in terms of technical and computational
effort. For this reason, the anatomical data often need to be complemented
with statistical estimates in order to define complete network models
of the brain. Filling in the gaps in the known connectivity in this
way may be referred to as \textit{predictive connectomics}. The corresponding
predictions have to be validated in some way, for instance by leaving
out part of the known anatomical data and determining how well these
are reproduced by the statistical estimates.

Understanding the human brain is often considered the holy grail of
neuroscience, not least because of the hope of finding novel cures
and therapies for brain diseases. However, due to its size and enormous
complexity, it can be helpful on the way to this goal to investigate
simpler, more tractable brains of other species. Eric Kandel took
this approach in his famous studies on the sea slug \textit{Aplysia}
\cite{Kandel07_search}, and it is a guiding thought behind the OpenWorm
project on modeling \textit{C. Elegans}. Furthermore, data obtained
with invasive methods are, for obvious reasons, much more abundant
for non-human brains. Of course, understanding the brains of species
besides humans can be seen as a valuable aim in itself---for improving
the well-being of animals, for inspiring industrial applications,
or as an intellectual pursuit, like cosmology or paleontology, which
enriches us culturally even if it has no direct practical application.
And, as it is with all basic sciences, one never knows what innovations
the knowledge gained may inspire many years into the future. For these
reasons, we do not restrict ourselves to the human brain, but also
consider various other species. However, we focus on mammalian brains,
which exhibit qualitative similarity to the human brain and may therefore
teach us most about our own brains. Non-human primate brains deserve
particular attention, as they are closest to the human brain in terms
of anatomy and function. Although extensive differences in detailed
organization remain \cite{Orban04_315,Sereno05_135,Hutchison12_29}
the anatomical similarities and evolutionary path give hope that universal
principles can be discovered extending to the human brain. Furthermore,
the chapter has an emphasis on our study object of choice---the cerebral
cortex.

To limit the scope of the chapter, we also restrict ourselves to anatomical
properties relevant for networks of point neurons or neural populations,
neglecting most aspects of detailed neuron morphology and placement
of synapses on the dendritic tree and axonal arborizations. The anatomical
characteristics entering into the definition of such neural network
models can be classified into brain morphology, cytoarchitecture,
and structural connectivity. Brain morphology describes geometric
macroanatomical properties, for instance the thickness of the cerebral
cortex and its layers, or the curvature.\textit{ }Cytoarchitecture
refers to the composition of brain regions in terms of the sizes,
shapes, and densities of neurons. Structural connectivity\textit{
}refers to properties of the synaptic connections between neurons,
including numbers of synapses between a given pair of neurons, or
the probability for neurons from two given populations to be connected.

The type and level of detail of anatomical information that is required
depends on the type and aim of the modeling study. A population model,
describing only the aggregate activity of entire populations of neurons,
does not require the connectivity to be resolved at the level of individual
neurons, nor is it generally necessary to know the number of neurons
in each population for such models. For models resolving individual
neurons, in some cases it may be of interest to incorporate detailed
connectivity patterns, while sometimes population-level connection
probabilities suffice. The difference lies in the questions that the
different types of models allow one to address. In one approach, the
modeler tries to derive as realistic a connectivity matrix as possible,
in the hope of obtaining the best possible predictions of dynamics
and information processing on the anatomical substrate. Here, it always
needs to be kept in mind that more detail does not necessarily mean
better predictions: adding more parameters can actually reduce the
predictive power of a model, for instance when these parameters are
not sufficiently constrained \cite{Jolivet08_417-426,Teeter18_709}.
However, if this approach is successful, it in principle allows the
effects of detailed physiological parameter changes on network dynamics
to be predicted (somewhat akin to weather forecasts), which may ultimately
find clinical applications. In a contrasting modeling approach, connectivity
features are abstracted and the influence of these abstract features
(e.g., small-worldness, clustering, hierarchical organization, etc.)
on graph theoretical, dynamical, or functional properties of the network
are investigated. This approach places less emphasis on strict biological
realism and attempts to provide a more conceptual understanding of
the links between brain anatomy, dynamics, and function. In practice
there is a continuum of approaches between these two extremes. For
instance, models may incorporate biologically realistic features at
an intermediate level of detail (e.g., population-specific connection
probabilities without detailed connectivity at the single-neuron level)
in order to simultaneously enable conceptual scientific conclusions
and a degree of validation of these conclusions by direct model comparisons
with experimental data.

Formulating and parametrizing neuronal network models is still often
a painstaking effort, where the researcher digs through a vast literature
to collect the relevant parameter values, from disparate experimental
methods and labs. This systematization of the available knowledge
into a common framework forms a central part of computational modeling
work, and allows future researchers to continue at the next level
of complexity. It is also highly specific to the modeling problem
and data modalities at hand, so that we cannot give one-size-fits-all
advice on how to deal with and interpret anatomical data to develop
network models. However, we can provide general guidance regarding
what to look out for in the various data modalities, and how to incorporate
the corresponding data into models. Furthermore, data are increasingly
collected in systematic databases, which make the modeler's life easier
by offering comprehensive data obtained with the same experimental
methods, often even from the same lab. Most promising for facilitating
this process are recent multilevel brain atlases, which aggregate
both macro- and microstructural information into systematic anatomical
reference frameworks.

In this chapter, we provide an overview of the types of anatomical
information that can be used to define biological neural network models,
point to available resources and databases, and describe methods
for predicting connectivity and validating the predictions. The text
considers physiological properties only where they relate directly
to anatomy. This overview is intended as an aid for computational
neuroscientists to develop accurate models of biological neuronal
networks. 

\section{Brain morphology and cytoarchitecture}

In this section, we describe the main types of information on the
morphology and cytoarchitecture of brain regions, and corresponding
resources available to modelers. We start by providing a brief introduction
to brain atlases, which systematize information on these anatomical
properties. Next, we treat the morphological property of cortical
and laminar thicknesses in more detail. We then go into the determination
of neural population sizes and the location of neurons within brain
regions, and close with a short discussion of the use of morphology
and cytoarchitecture in computational models. We do not distinguish
between cell types within regions, as this would substantially extend
the scope of the chapter, and, especially in the context of network
models that do not resolve neural compartments, more directly concerns
chemical and electrophysiological instead of anatomical properties.

\subsection{Brain atlases\label{subsec:Brain-atlases}}

Brain atlases\textit{ }are a tool for defining brain areas and aggregating
regional descriptions of the brain in a consistent anatomical framework.
A brain atlas typically consists of a template space, a set of maps
or a parcellation, and a taxonomy,\emph{ }which provides the names
and mutual relationships of those regions.

The template space of a brain atlas is typically represented by one
or multiple scans of a brain, which provide an anatomical description
of an underlying standardized coordinate space\emph{. }Depending on
the task at hand, different template spaces are used. A classical
template space for the human brain is Talairach
space \cite{Talairach88}, which assumes that the relative distances
between brain regions are preserved between individuals, and defines
a rescalable grid accordingly. Talairach coordinates are still in
wide use in functional neuroimaging. Today, it is more common to use
one of the MNI templates defined by the Montreal Neurological Institute
\cite{Lancaster07_1194,Laird10_677}, which include single- and multi-subject
averages of MRI scans as volumetric standard spaces. While the MNI
templates define standard spaces at millimeter resolution, the BigBrain
offers a brain model of a single subject based on a three-dimensional
reconstruction from $7,\!400$ histological sections, at an isotropic
resolution of $20\:\mu\mathrm{m}$ \cite{Amunts13_1472}. As the tissue
sections were stained for cell bodies, this model provides the most
detailed three-dimensional reference of human cytoarchitecture available
today. Ongoing research addresses the three-dimensional cellular-level
reconstruction of brains at $1\:\mu\mathrm{m}$ resolution, which
poses considerable techical challenges for human brains due to their
size and topological complexity \cite{Dickscheid19_223}.

Brain maps and parcellations assign brain regions to coordinates of
a template space. In case of a standard whole-brain parcellation,
each voxel has a unique region index, and the assigned regions do
not overlap. In case of probabilistic maps, however, each coordinate
is assigned a probability to belong to any of the regions, resulting
in a set of overlapping maps to define the atlas.  Parcellations
are based on different modalities\textit{ }\textit{\emph{of brain
organization, including cytoarchitecture}}\textit{ }(e.g. \cite{Amunts2020_onlinefirst}),
chemoarchitecture\emph{ }(spatial distribution patterns of molecules
like specific neurotransmitter receptors, e.g. \cite{Zilles04}),
structural connectivity (patterns of connectivity with other brain
regions as defined by axonal connections, e.g. \cite{Eickhoff15_4771,Fan16_3508}),
functional connectivity (spatial co-activation patterns under different
cognitive conditions (e.g. \cite{Gordon16_288}), anatomical landmarks,
or a combination of such features in the case of multimodal parcellations
\cite{Bohland09_9,Arslan18_5,VanEssen18_640}. 

The gold standard of brain parcellations is based on cytoarchitecture
as measured in histological sections. The early Brodmann atlas of
the cerebral cortex of humans and other primates uses such a cytoarchitectonic
parcellation \cite{Brodmann1909}. Some years later, von Economo and
Koskinas developed an atlas \cite{vonEconomo1925} with a more comprehensive
characterization of the cortical layers, and taking into account cortical
folding by describing cytoarchitecture orthogonal to the cortical
surface. However, the bases of these pioneering works remain collections
of separate brain slices, thereby lacking coverage of the full three-dimensional
anatomical space, as well as of the variability across subjects. Recent
work in probabilistic cytoarchitectonic mapping addresses the latter
challenge by aggregating microscopic maps from ten different subjects
in MNI space \cite{Amunts2020_onlinefirst}. Furthermore, different
groups are working on full three-dimensional, microscopic resolution
maps of cytoarchitectonic areas \cite{Schiffer19_OHBM} and cortical
layers \cite{Wagstyl18_2551} in the BigBrain model, giving access
to region- and layer specific measures of, e.g., cell densities and
laminar thickness.

In connectivity-based parcellation, voxels with similar connection
properties are grouped together \cite{Eickhoff15_4771}. An example
of an atlas using connectivity-based parcellation is the human Brainnetome
Atlas \cite{Fan16_3508}, which takes the Desikan-Killiany atlas based
on cortical folds (the sulci and gyri) \cite{Desikan06_968} as its
starting point. The Brainnetome atlas has the advantage for modeling
studies that data on functional connectivity, a term used in neuroscience
for activity correlations, is freely available in the same parcellation,
allowing straightforward testing of model predictions on network dynamics.

The Allen Institute has published multiatlases of the developing\footnote{BrainSpan Atlas of the Developing Human Brain (2011) http://brainspan.org.
Funded by ARRA Awards 1RC2MH089921-01, 1RC2MH090047-01, and 1RC2MH089929-01.} and adult human brain \cite{Shen12_711,Sunkin12_D996}, mapping cytoarchitecture,
gene expression, and for the adult brain also connectivity as measured
with diffusion tensor imaging (DTI), a magnetic resonance imaging
method that detects axon tracts. This multimodality, where different
types of data are represented in the same template space and parcellation,
is useful for modelers, not only because of the richness of the data,
but also as mapping data from different sources between template spaces
and parcellations introduces inevitable errors. 

The macaque, as a close relative of humans, is an important model
organism, for which several atlases have been created. These include
the atlas of Markov et al. (2014) \cite{Markov14} with the so-called
M132 parcellation of 91 cortical areas, and a whole-brain atlas by
Calabrese et al. (2015) \cite{Calabrese15_408} based on DTI. Another
commonly studied species is the mouse, for which state-of-the-art
atlases of gene expression data \cite{Lein07_168}, cytoarchitecture
as measured with Nissl staining,\textit{ }which stains nucleic acids
and thereby cell bodies of both neurons and glia\textit{, }and mesoscopic
connectivity obtained by anterograde viral tracing \cite{Dong08_Allen,Kuan15_4}
are provided by the Allen Institute. Paxinos and Franklin provide
the other most commonly used mouse brain atlas \cite{Paxinos2019},
which recent work combines with the Allen Institute coordinate framework
\cite{Chon19_1}.

Several online resources exist for browsing brain atlases. The Scalable
Brain Atlas provides web-based access to a collection of atlases for
the human brain and for a number of other mammals, including macaque,
mouse, and rat \cite{Bakker15}. The Human Brain Project provides
online services for interactive exploration of atlases for the mouse,
rat, and human brain through the EBRAINS infrastructure\footnote{https://ebrains.eu/services/atlases/brain-atlases}.
The human brain atlas is a multilevel framework based on probabilistic
atlases of human cytoarchitecture, and includes links with maps of
fiber bundles and functional activity, as well as a representation
of the microscopic scale in the form of the BigBrain model with maps
of cortical layers and cytoarchitectonic maps at full microscopic
resolution \cite{Schiffer19_dataset}.

\subsection{Cortical and laminar thicknesses}

The geometrical properties of the global and regional morphology of
the brain have obvious relevance for brain models that explicitly
represent space, but can also be important for estimating connectivity
and numbers of neurons in non-spatial models. These properties include
coordinates of region boundaries, spatial extents of brain regions,
and properties of regional substructures such as thicknesses of cortical
layers. Coordinates and spatial extents of brain regions are captured
by atlases as described in the previous section. Another geometric
property that is often of interest is the thickness of cortex and
its layers.

Cortical and laminar thicknesses can be either determined directly
from histology of brain slices, or using structural MRI. When the
MRI scans have sufficiently high resolution, these methods yield comparable
results \cite{Fischl00_11050,Luesebrink13_122,Cardinale14_535,Wagstyl18_35},
but both methods have their own drawbacks. Brain slices generally
represent sparse samples, are difficult to obtain precisely perpendicularly
to the cortical sheet, and are subject to shrinkage, which has to
be controlled for. Furthermore, identification of layers and the boundary
between gray and white matter is still often performed manually, although
automatic procedures are under development \cite{Wagstyl18_2551,Li2019_1}.
Structural MRI can cover the entire cortex and at least the gray/white
matter boundary tends to be segmented using computer algorithms, but
it has a lower resolution in the section plane than microscopy of
brain slices, the exact resolution depending on the strength of the
scanner and the scanning protocol. Von Economo provides laminar and
total cortical thicknesses for all areas of human cortex based on
$25\:\mu\mathrm{m}$ sections \cite{VonEconomo09}. More recently,
cortical and laminar thicknesses (the thicknesses of the individual
cortical layers) have been identified in the BigBrain, forming a state-of-the-art,
comprehensive dataset on human cortex \cite{Wagstyl18_2551,Wagstyl2020_e3000678}.
The gray and white matter volumes and surfaces, along with the layer
surfaces, are freely available\footnote{ftp://bigbrain.loris.ca/}
and can be explored interactively in the EBRAINS human brain atlas
viewer. Alvarez et al. (2019) \cite{Alvarez19_116057} determined
the thicknesses of $25$ human visual areas from $700\:\mu\mathrm{m}$
resolution MRI data from the Human Connectome Project, also making
the quantitative area-averaged data freely available. Calabrese et
al. (2015) \cite{Calabrese15_408} derived macaque cortical thicknesses
from MRI scans at $75\:\mu\mathrm{m}$ resolution, available as an
image file. Hilgetag et al. (2016) \cite{Hilgetag16} provide total
cortical thicknesses for $22$ vision-related cortical areas of the
macaque monkey, determined from brain slices sampled every $150-200\:\mu\mathrm{m}$
throughout the region of interest. At least in the vision-related
areas of macaque cortex, total cortical thickness correlates inversely
with neuron density, so that a statistical fit allows the thicknesses
of the remaining vision-related areas to be estimated \cite{Schmidt18_1409}.
Correspondingly, cortical thickness varies systematically along the
anterior-posterior axis in primates \cite{Cahalane12}. Rough estimates
of the laminar thicknesses of macaque vision-related areas based on
a survey of micrographs (microscopic images) have been published \cite{Schmidt18_1409}.
Comprehensive data on cortical thicknesses of other species are sparse,
especially in a form that is directly usable by modelers. Methods
for extracting cortical thicknesses from MRI in rodents are under
development \cite{Pagani16_62,Feo19_82}.

\subsection{Numbers of neurons}

Another basic property of brain circuits is their numbers of neurons,
which can be determined from the size of brain regions and their neuron
density. Over the years, different methods of counting cells have
been used \cite{West93_275,Miller14_36}. When total cell counts are
of interest and their precise distribution across space is less important,
tissue can simply be homogenized and the numbers of cell nuclei suspended
in a fluid can be counted in samples under a microscope. The isotropic
fractionator is a version of such a homogenization and direct counting
method\textit{ }\cite{Herculano05_2518}\textit{. }The term `fractionator'\textit{
}refers to a uniform random sampling scheme which divides samples
into `fractions' or counting boxes, enabling a statistical estimate
of total cell counts to be obtained by considering only some fractions
\cite{West91_482}.

Stereological methods are a more involved class of methods that determine
three-dimensional properties from two-dimensional sections through
the tissue. The advantage of these methods is that the cells are counted
in their real three-dimensional environment (depending on the section
thickness) and thus spatial and area-specific values can be collected,
e.g. cell densities in a single cortical lamina. Beside the fact that
most stereological methods are quite labor- and time-intensive, the
problem arises that the same cell may appear in two or more sections
but should only be counted once. The disector addresses this issue
by considering pairs of adjacent sections and only counting the cells
that are present in the second but not the first section, effectively
counting only the `tops' \cite{Sterio84_127}. The success of this
approach depends on being able to recognize if features in the adjacent
sections belong to the same cell, and on effectively correcting for
large structures that extend across more than two sections. The optical
fractionator combines the aforementioned uniform sampling method (the
`fractionator') with optical disection, in which objective lenses
with a high numerical aperture are used to focus through the tissue
to identify individual cells. A guarding zone above and below the
inspected volume prevents multiple counting of truncated structures.

For cell bodies to be identified under the microscope, they are first
dyed. Two commonly used methods are the aforementioned Nissl staining,
and antibody staining of the protein NeuN that is present in the nuclei
of most vertebrate neurons but not in glia \cite{Mullen92_201}. Another
technique dying both neurons and glia is silver staining \cite{Merker83_235},
used for instance in the BigBrain model.

\begin{figure}
\includegraphics[width=1\textwidth]{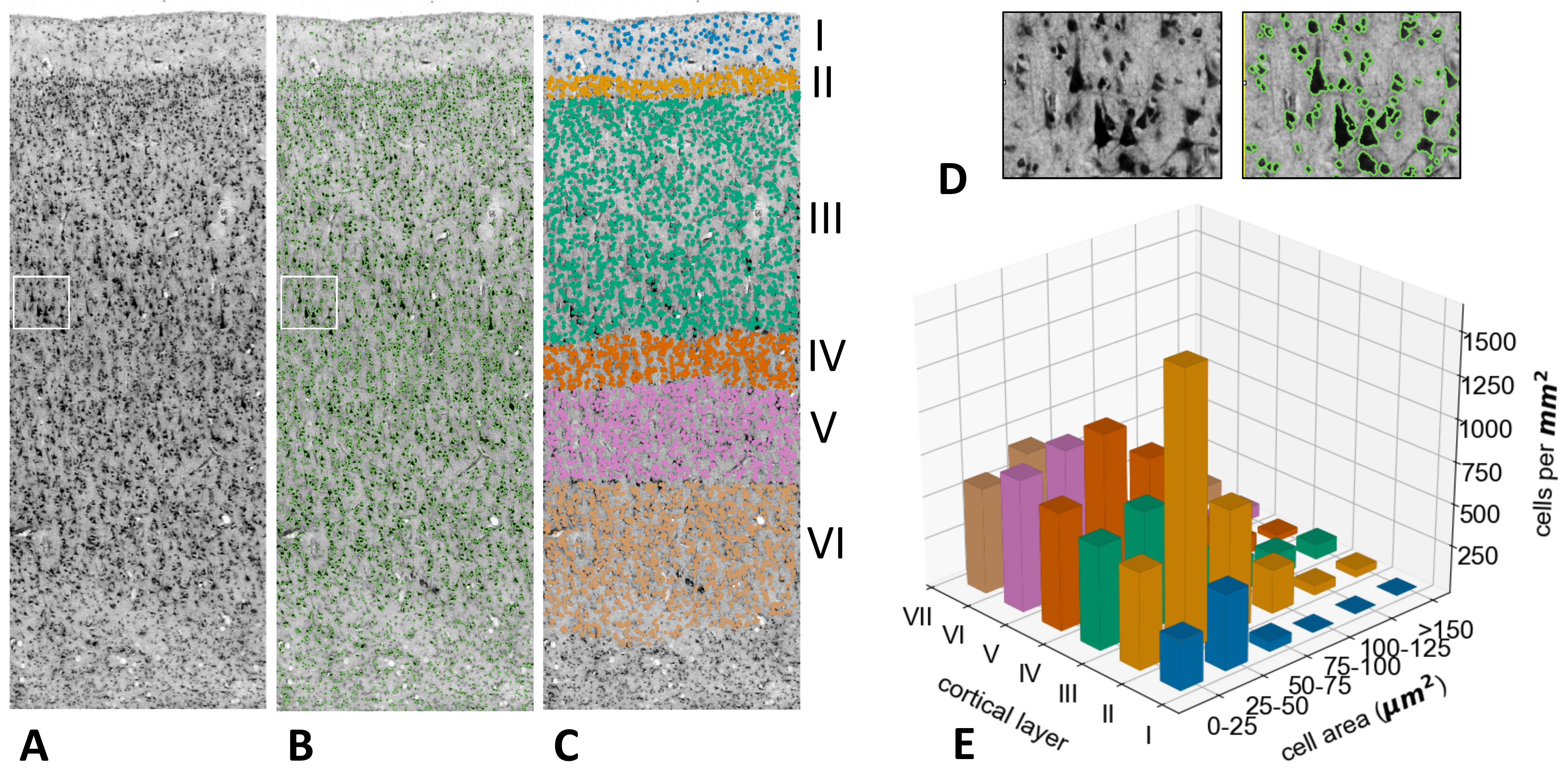}

\caption{Extraction of layer-specific cell density estimates from microscopic
scans of histological sections stained for cell bodies. \textbf{A.}
Cortical patch of a scan. \textbf{B.} Example result of automatic
instance segmentation of cell bodies using state-of-the-art image
analysis (E. Upschulte, Institute of Neuroscience and Medicine, Forschungszentrum
Jülich). \textbf{C.} Centroids of detected cell bodies, colored by
cortical layer. \textbf{D.} Zoom into the local region of interest
indicated by the white rectangle in Panels A and B. \textbf{E.} Two-dimensional
histogram showing the number of cells in each layer, grouped by area
of the cell body as segmented in the image.}
\label{fig:cell_densities}
\end{figure}

A number of comprehensive data sets on cell and neuron counts are
available, although estimates can vary quite a bit across studies
\cite{Ero18_84}. Overall numbers of neuronal and non-neuronal cells
have been estimated for the brain as a whole, and for its major components
like the cerebral cortex and the cerebellum, for a large number of
species\footnote{https://en.wikipedia.org/wiki/List\_of\_animals\_by\_number\_of\_neurons}
\cite{Herculano06_12138,Azevedo09_532,Herculano-Houzel09,Sarko09_8,Herculano2012_10661}.
In most cases, these cell numbers were acquired using the above-described
techniques based on homogenized tissue. The von Economo atlas contains
cell densities for human cortex with areal and laminar resolution,
as determined with Nissl staining \cite{VonEconomo09}. Because the
Nissl technique stains both neurons and glia, which can, however,
be distinguished based on morphology, it is not entirely clear whether
glia are included in these cell densities. Furthermore, the cell numbers
were measured without modern stereological approaches and without
characterizing inter-individual variability. Modern high-performance
computing methods are being applied for image registration of two-dimensional
cortical and subcortical images to determine three-dimensional cell
distributions \cite{Dickscheid19_223} (\figref{cell_densities}),
laying the foundation for future quantitative data sets representing
an update and refinement with respect to the von Economo study. Collins
et al. \cite{Collins10_15927} provide cortical area-specific neuron
densities for the non-human primates galago, owl monkey, macaque,
and baboon as determined with the isotropic fractionator. So-called
cortical types\textit{ }or architectural types characterize the neuron
density and laminar differentiation of primate cortical areas in a
discretized manner, and thereby enable rough neuron density estimates
where these have not been directly measured \cite{Barbas97,Dombrowski01,Schmidt18_1409,Garcia-Cabezas2019_985,Hilgetag19_905}.
Herculano-Houzel et al. (2013) \cite{Herculano-Houzel13_35} measured
neuron and cell counts and densities for the areas of mouse isocortex.
Keller et al. (2018) \cite{Keller18_83} systematically reviewed region-specific
neuron and glial densities throughout the mouse brain. Structures
that have been characterized in detail also include the somatosensory
areas of rat cortex and thalamus \cite{Meyer2010_2277,Markram2015_456}.
Despite many more data having been published, a large number of species-specific
brain region compositions are still unknown, especially for subcortical
regions. Scaling laws across species enable numbers of neurons to
be estimated based on structural properties like brain and regional
mass and volume \cite{Braitenberg01,Herculano06_12138,Azevedo09_532,Herculano-Houzel09,Sarko09_8,Herculano2012_10661}.

Neuron counts or densities may not always be available in the particular
parcellation chosen by the modeler. A mapping between parcellations
may be performed by determining the overlaps between areas in different
parcellations, for which the parcellations have to be in the same
reference space. A large number of methods for registering images
to the same reference space using nonlinear deformations have been
developed \cite{Dale99_179,Fischl99_195,Thompson02_13}. For macaque
atlases registered to the so-called F99 surface, a tool provided alongside
the CoCoMac database on macaque brain connectivity\footnote{http://cocomac.g-node.org/services/f99\_region\_overlap.php}
\cite{Stephan01_1159} calculates the absolute and relative overlaps
between cortical areas. The data in the new parcellation can then
be computed as a weighted sum over the contributions from the areas
in the original parcellation. However, this method entails the assumption
that the anatomical data for each given area are representative of
that area as a whole, and neglects inhomogeneities within areas. It
should further be noted that criteria for area definitions, such as
their cytoarchitecture or connectivity, are likely to provide information
beyond this purely spatial approach. Nonlinear
image registration techniques can take such factors into account,
or alternatively, a coordinate-independent mapping can be performed
\cite{Stephan00_37}. No perfect solution for mapping anatomical data
between parcellations exists, but in general, the more criteria are
considered, the better the mapping.

\subsection{Local variations in cytoarchitecture}

Even within brain regions, cell densities are not constant but display
local variations. An example of known spatial organization of neuron
positions are so-called cortical minicolumns, also known as microcolumns,
arrangements of on the order of $100$ neurons perpendicular to the
cortical surface, across the cortical layers. Cortical macrocolumns
or hypercolumns are millimeter-scale structures containing thousands
or tens of thousands of neurons with similar response properties in
one or a few coding dimensions, for instance ocular dominance or position
in the visual field. Cortical macrocolumns are particularly pronounced
in the barrel cortex of rodents, which encodes whisker movements.
In barrel cortex, the `barrels' are cylindrical structures in layer
IV containing neurons that respond preferentially to a particular
whisker and have response properties and connectivity distinct from
the interbarrel regions.

Various data on variations in neuron density within brain regions
are available. Probably the most comprehensive data set of three-dimensional
cell distributions is the Allen Mouse Brain Atlas, which contains
both neurons and glia \cite{Ero18_84}. Spatial gradients in retinal
cell densities have been well characterized \cite{Drager81_285,Stone81_231,Curcio90_5,Waessle94_561,Euler95_461,Shand00_176},
and those in thalamus to a lesser extent (e.g., \cite{Ahmad93_631}).
The vertical distribution of cells in several cortical areas has also
been characterized at a spatial resolution beyond that of cortical
layers \cite{Mitra55_467,Sloper79_141,Cozzi17_2743}.

Studies resolving small cortical patches provide a sense of the variability
of neuron density across the cortical sheet within primate cortical
areas \cite{Collins10_15927,Turner16_1}. Furthermore, many studies
have subdivided brain regions into discrete components with different
cellular compositions, e.g., \cite{McDonald82_401,Stepniewska97_1043,Voogd98_307,Duvernoy05}.

\subsection{Use of morphology and cytoarchitecture in models}

While most neural network models specify their architecture using
concepts such as areas and layers, in some cases the neurons are simply
assigned positions in continuous three-dimensional space and the connectivity
is specified without reference to such concepts (e.g. \cite{Schumann2017_160}).
In the conceptual approach, different connectomes may be obtained
depending on the chosen parcellation. The particular choice of parcellation
for instance affects topological properties of the corresponding connectomes
\cite{Romero12_3522,deReus13_397}. Apart from this `gerrymandering'
issue, when predictive connectomics is used to fill in gaps in connectivity
data with the conceptual approach, the choice of parcellation may
influence the results. The findings of \cite{Romero12_3522,deReus13_397}
for instance imply that incomplete connectomes completed via topological
rules could differ depending on the parcellation. In view of the variability
induced by differences between parcellations, there is something to
be said for the continuum approach when the data allow it. Interpretation
of the network dynamics in terms of region-specific activity may then
be done in a post-hoc manner, flexibly with regard to the region definitions.

In spatially extended models, the neurons may be placed on a regular
grid, with some jittering, at random positions, or at precise coordinates
in space. Here, artificial symmetries in the network dynamics due
to grid-like placement of neurons, which may arise for instance when
the connectivity and delays are directly determined by the distances
between neurons, should be avoided. Besides informing connectivity,
the positions can be important for predicting signals with spatial
dependence, like the local field potential (LFP), electroencephalogram
(EEG) or magnetoencephalogram (MEG).

Precise region shapes are so far hardly used in computational modeling.
Rather, the relatively rare network models that take into account
three-dimensional structure tend to restrict themselves to simple
geometric shapes like cubes or cylinders. An available but not yet
widely used tool enables three-dimensional region volumes to be modeled
through a combination of deformable two-dimensional sheets, where
atlas data or histological images can support the modeling process
via integration with the software Blender \cite{Pyka14_91}. In an
example application, the three-dimensional shape of the hippocampus
was shown to substantially affect the connectivity between neurons
predicted based on their distance. Accurate representations of volume
transmission effects such as ephaptic coupling (non-synaptic communication
via electrical fields or ions) \cite{Anastassiou15_95}, as well as
the prediction of meso- and macroscopic signals like the LFP, EEG,
and MEG also rely on the spatial distribution of neurons and thus
benefit from measured three-dimensional brain morphology \cite{Jirsa01_286,Hagen16,Hagen18_92}.

On the scale of local microcircuits on the order of a millimeter,
spatial variations in cortical and laminar thicknesses across the
cortical sheet within each area are limited and are generally ignored
in computational models. Cortical and laminar thicknesses are then
straightforwardly incorporated by scaling the numbers of neurons accordingly,
and sometimes by distributing the neurons across cortical depth. In
future, as resources become available for modeling extended cortical
regions in detail, continuous variations in cortical and laminar thicknesses
may be incorporated.

It is also not yet common for computational models to take into account
continuous variations in neuron density within brain regions. However,
a number of models already divide regions into discrete subdivisions
with different cellular compositions, e.g., \cite{Casali19_37}. The
organization of cortex into minicolumns and macrocolums has been incorporated
for instance in models of attractor memory \cite{Lundqvist06_253,Johansson07_1871}
motivated by a functional interpretation. In future, increasingly
realistic placement of neurons in models may yield more sophisticated
predictions of spatially resolved brain signals and of network dynamics,
through associated properties like distance-dependent connectivity.

\section{Structural connectivity}

Neurons in the brain exchange chemical signals via synapses, and in
some cases are in more direct contact via so-called gap junctions.
Although gap junctions are probably important for some phenomena (e.g.
\cite{Traub01_9478}), we here focus on the former, much more numerous
type of connections, the synapses. The huge number of synapses in
mammalian brains has so far precluded mapping all of them individually,
although efforts are underway towards dense reconstruction of the
mouse brain \cite{DeWeerdt2019_S6}. However, various methods exist
for measuring neuronal connectivity, at scales ranging from individual
synapses to entire axon bundles between areas. While some models distinguish
individual synapses and thus need information at this level, other
models lump synapses together, so that aggregated connectivity information
suffices.

This section provides an overview over available types of information
on neuronal network connectivity, along with resources and databases
that can be used for constructing neuronal network models. We describe
connectivity information according to the major experimental methods:
microscopy, paired recordings, glutamate uncaging, axonal tracing,
and diffusion magnetic resonance imaging (diffusion MRI), of which
the most commonly used form is diffusion tensor imaging (DTI).

\subsection{Microscopy\label{subsec:Microscopy}}

The oldest and lowest-resolution form of microscopy is light microscopy,
providing a magnification factor of up to about $1,\!000$. Neuron
reconstructions from light microscopy of adjacent tissue slices allow
rough estimates of connectivity based on the proximity of pre- and
postsynaptic neural processes (cf. \subsecref{Peters'-rule}). Following
this approach, Binzegger et al. (2004) \cite{Binzegger04} derived
a population-level local connectivity map for cat primary visual cortex.
However, as detailed in \subsecref{Peters'-rule}, predicting connectivity
based on proximity has its drawbacks, which should be kept in mind
when interpreting the resulting connectomes. Furthermore, tissue slicing
cuts off dendrites and axons, which may extend over millimeters and
more, so that assessing medium- to long-range connectivity requires
extensive three-dimensional reconstructions. A method that facilitates
such reconstructions is block-face tomography, in which scanning of
the surface of a tissue block is alternated with the removal of thin
slices from the surface \cite{Denk04_11}.

Two-photon microscopy is a sub-micron resolution imaging technique
that uses laser irradiation of tissue to elicit fluorescence through
two-photon excitation of molecules \cite{Denk90_73}. A high-throughput
block-face tomography pipeline has enabled the reconstruction of the
full morphologies of $1,\!000$ projection neurons in the mouse brain
at a resolution of $0.3\times0.3\times1\:\mu\mathrm{m^{3}}$, the
MouseLight data set of Janelia Research Campus \cite{Economo16_e10566,Winnubst19_268}.
A viewer for the MouseLight morphologies is available\footnote{https://neuroinformatics.nl/HBP/mouselight-viewer/}.
A finding that stands out from this data set is the remarkable variability
in projection patterns, each neuron projecting to a different subset
of target regions for the given source region.

At nanometer spatial scales, electron microscopy enables the identification
of individual synapses and the precise shape and size of the presynaptic
and postsynaptic elements, even down to individual synaptic vesicles.
This method is extremely labor-intensive, but heroic efforts have
nevertheless led for instance to estimates of synapse density in different
areas of human cortex \cite{Alonso08_14615,Alonso11_8} , a volume
reconstruction of the entire Drosophila (fruit fly) brain \cite{Zheng18_730},
the morphological reconstruction of $1,\!009$ neurons in a microcircuit
of rat somatosensory cortex \cite{Markram2015_456}, and full reconstructions
of $1,\!500\:\mu\mathrm{m^{3}}$ \cite{Kasthuri2015_648} and more
recently $>5\times10^{5}\:\mu\mathrm{m^{3}}$ \cite{Motta2019_eaay3134}
of mouse cortical tissue. A noteworthy finding from these studies
is that the presence of synapses is not perfectly determined by the
close proximity of axons and dendrites (appositions). For instance,
an apposition is far more likely to predict an actual synaptic contact
for pairs of neurons that also form synapses elsewhere on the axon
and dendrite \cite{Kasthuri2015_648}. Such a rule will tend to lead
to a long-tailed distribution of the multiplicity of synapses between
pairs of neurons.

Synapses may look asymmetric or symmetric under the microscope, where
asymmetric synapses have a pronounced postsynaptic density and are
predominantly excitatory, while symmetric synapses have roughly equally
thick pre- and postsynaptic densities and tend to be inhibitory. Both
the size of synapses and their location on dendrites are informative
about their effective strength in terms of postsynaptic potentials
evoked at the soma \cite{Spruston94,Murthy01_673,Harris03_745,Kwon17_1100}.
Furthermore, synapse locations on dendrites can tell us something
about their interaction with other synapses; however, these complex
interactions are not captured by point neuron or population models.
Axonal varicosities\textit{ }or boutons are swellings along axons
(boutons en passant) or at axon terminals (terminal boutons) that
host synapses, and which are detectable through all microscopic methods
mentioned here. Even when the synapses themselves are not directly
imaged, boutons may be taken as evidence for synapses, with the caveats
that some synapses are not established on boutons, and individual
boutons may contain different numbers of synapses \cite{Rodriguez20_2663}.

In summary, microscopy is useful for estimating connectivity based
on appositions, reliable estimates of numbers of synapses in a given
volume, detailed connectivity features such as the multiplicity of
synapses between pairs of neurons, and correlative information on
synaptic efficacy.

\subsection{Paired recordings \label{subsec:Paired-recordings}}

In paired recordings, electrodes are used to simultaneously stimulate
one cell and measure the response in another cell, either \textit{in
vitro} or \textit{in vivo}. Stimulation may be performed extracellularly,
intracellularly with sharp electrodes, or via patch clamp; recordings
normally use one of the latter two techniques. This method sums up
the contributions from potentially multiple synapses between the pair
of neurons, which should be kept in mind when incorporating the corresponding
synaptic strengths into models. Where anatomy-based methods can
have the drawback that they do not provide conclusive evidence for
physiologically active synapses, paired recordings identify functional
synapses. However, existing connections may be missed depending on
the experimental protocol, for instance due to axons and dendrites
being cut off during slice preparation. Each pair of neurons should
also be tested multiple times, because in individual trials, axonal
or synaptic transmission failures may occur, or the postsynaptic potential
may be too small to be detectable among the noise \cite{Debanne08_1559}.
Paired recordings may be biased toward neurons that are easier to
patch or insert an electrode into, for instance larger cells. Especially
\textit{in vivo}, where the network exhibits background activity,
responses may in principle be caused by activation of neurons other
than the one that is stimulated. Responses are judged to be monosynaptic
based on a short, consistent response latency, usually of a few tenths
of milliseconds \cite{Berry76_1,Sedigh17_5250}.

Most paired recordings are highly local, with a distance no greater
than $100\:\mu\mathrm{m}$ between the somas of the pre- and postsynaptic
cells. They provide the modeler with connection probabilities in terms
of the fraction of pairs of neurons that have at least one synapse
between them. For interpreting these connection probabilities, it
is important to take into account the spatial range of the recordings,
as connection probability is generally distance-dependent. The measurements
represent a spatial average over this distance-dependent connectivity,
which is in mathematical terms a double sum (which may in continuum
approximation be represented by an integral) over the positions of
the source and target neurons.

Paired recordings show that, on the scale of local microcircuits up
to $200\:\mu\mathrm{m}$ from the presynaptic soma, bidirectional
connections between pyramidal neurons in cortical layer V occur significantly
more often than would be expected by chance \cite{Markram97b}. In
some studies, researchers have recorded from multiple neurons simultaneously
\cite{Thomson02_936,Song05_0507,Perin11,Kodandaramaiah18_e24656}.
Simultaneous recordings from respectively four \cite{Song05_0507}
and twelve \cite{Perin11} rat cortical neurons confirm the overrepresentation
of bidirectional connections regardless of the distance from the soma.
This type of analysis has also revealed that motifs with clustered
connections among three or more neurons are more common in the cerebral
cortex than would be predicted based on pairwise connection probabilities
alone \cite{Song05_0507,Perin11} (cf. \subsecref{Connectome-topology}).

\subsection{Glutamate uncaging\label{subsec:Glutamate-uncaging}}

Similarly to paired recordings, glutamate uncaging generates action
potentials in presynaptic neurons and records the response in postsynaptic
neurons connected to them. Usually, the method is applied to slice
preparations and neurons are recorded intracellularly, but \textit{in
vivo} application and extracellular recordings are also possible.
First, a compound consisting of glutamate bound to another molecule
is introduced, for instance by bathing a brain slice in a solution
with the caged glutamate. Then glutamate is released by photolysis
of the compound through focal light stimulation, causing action potentials
in neurons with their soma close to the stimulation site. Brain slices
are generally scanned systematically, generating for each given target
neuron a grid-like map of response amplitudes for each stimulated
location.

Originally, glutamate was uncaged using ultraviolet light \cite{Callaway93},
but due to light scattering and a large uncaging area, this stimulated
multiple neurons, making the results harder to interpret. Two-photon
stimulation, in which photolysis is triggered by the absorption of
two photons, enables individual neurons and even individual dendritic
spines to be stimulated \cite{Nikolenko07_943,Noguchi11_2447}. As
with paired recordings, an issue is that it cannot be known with certainty
whether the responses are monosynaptic or emerge due to sequential
activation of two or more neurons, but short-latency responses time-locked
to presynaptic action potentials in the absence of background activity
reliably indicate monosynaptic connections. Another issue is that
the uncaged glutamate may directly influence the recorded neuron,
so that stimulations that lead to short-latency responses with excessive
amplitudes have to be excluded from analysis. Furthermore, the same
caveats as for paired recordings apply with regard to distance dependence
of connectivity, and potential cutting of dendrites and axons during
slice preparation.

Purely based on glutamate uncaging response maps, it is not possible
to directly derive a neuron-level connectivity map, because it is
unknown how many different presynaptic neurons are activated across
stimulation sites. However, by combining glutamate uncaging with imaging
of the neurons, the connectivity between neurons can be determined
\cite{Nikolenko07_943}. In the absence of such direct imaging, the
number of source neurons eliciting a given glutamate uncaging response
can be estimated by dividing by the unitary synaptic strength (the
PSP or PSC size due to a single presynaptic neuron), if an independent
estimate for the latter is available. If one in addition makes an
assumption about the average number of sites from which a given presynaptic
neuron is activated, which depends on the resolution of the stimulation
grid, this yields an estimate of the number of neurons impinging on
a given postsynaptic cell. Typically, action potentials can be elicited
in a given neuron from a handful of sites \cite{Dantzker00,Schubert2003_2961}.
Finally, one can derive a connection probability by dividing by the
approximate number of neurons in the stimulated volume. Clearly, many
assumptions and approximations are involved in such derivations, so
that it is currently still difficult to reliably determine the connectivity
of neural network models from glutamate uncaging data. However, in
some cases, data obtained by this method are the best available for
a given brain region, in which case one may proceed via such assumptions
\cite{Hooks2011_e1000572}.

\subsection{Axonal tracing}

The technique of axonal or neuroanatomical tracing entails injecting
a tracer, which can be a molecule or virus, which is taken up by
neurons and transported toward cell bodies or axon terminals. In anterograde
tracing, the tracer is transported in the forward direction toward
the synapses, while in retrograde tracing, it is transported in the
backward direction from axons toward the cell bodies of the sending
neurons. In practice, most tracers are to some extent both anterograde
and retrograde, but one transport direction dominates \cite{Lanciego11_157}.
Detection of the tracer happens in one of multiple ways: the tracer
may itself be fluorescent, it may be radioactively tagged or conjugated
with a dye or enzymatically active probe, or it may be detected via
antibody binding \cite{Saleeba19_897}. Axonal tracing is generally
performed in the living brain, after which the animal is sacrificed
to detect where the tracer has ended up, but some substances also
enable tracing in postmortem tissue and therefore even in the human
brain, albeit over limited distances \cite{Galuske00_1946,Tardif01_1045,Seehaus13_442}.
The method is well suited to characterizing medium-to-long-range connections
such as those between cortical areas. A number of tracers, especially
certain viral tracers, are transneuronal, crossing synapses and tracking
polysynaptic pathways \cite{Kuypers90_71}. Furthermore, it is possible
to perform double or even triple labeling to visualize the participation
of neurons in two or more connection pathways \cite{Koebbert00_327}.
Double labeling with retrograde tracers for instance suggests that
the vast majority of cortico-cortical projection neurons in macaque
visual cortex send connections either in the feedforward direction
or in the feedback direction, not both, with respect to the hierarchy
of visual areas \cite{Markov14}.

Tracer injections typically cover a millimeter-scale area, so that
multiple axons are traced at the same time, not individual ones. Because
of the local spreading of the tracer, axonal tracing does not provide
reliable information about the region immediately surrounding the
injection site. An important drawback of the method is that only up
to a few injections can be performed in each animal, so that data
have to be combined across many animals to obtain a complete connectivity
graph. This introduces inevitable inaccuracies due to inter-individual
differences. Because tracers are taken up by neurons indiscriminately,
conventional tracing does not allow the specific connections of separate
subpopulations of neurons to be identified, let alone of individual
neurons. However, over the past decades a number of viral tracing
methods have been developed that trace specific molecularly marked
neuronal subpopulations \cite{Saleeba19_897}. A modern technique
uniquely labeling neurons with random RNA sequences enables high-throughput
mapping of projections at the level of individual source neurons \cite{Chen19_772}.

While axonal tracing traditionally only gave qualitative information
about connectivity, for instance describing staining as sparse, moderate,
or dense, more recently a number of groups have gone through the painstaking
effort of counting the numbers of labeled cells in retrograde tracing
experiments. A notable quantitative tracing data set characterizes
the connectivity between a large number of areas in macaque cortex
in terms of overall fractions of labeled neurons (FLN) and fractions
of supragranular labeled neurons (SLN) in all source areas projecting
to each injected target area \cite{Markov14,Markov2014_17}. SLN relates
to the hierarchy of vision-related cortical areas, as feedforward
projections tend to emanate from layer II/III and thus have a high
SLN, while feedback projections emanate preferentially from infragranular
layers and have a low SLN. A similarly comprehensive resource of
quantitative retrograde tracing data is available for the marmoset
neocortex \cite{Majka16_2161,Majka20_1}.

\begin{figure}
\includegraphics[width=1\textwidth]{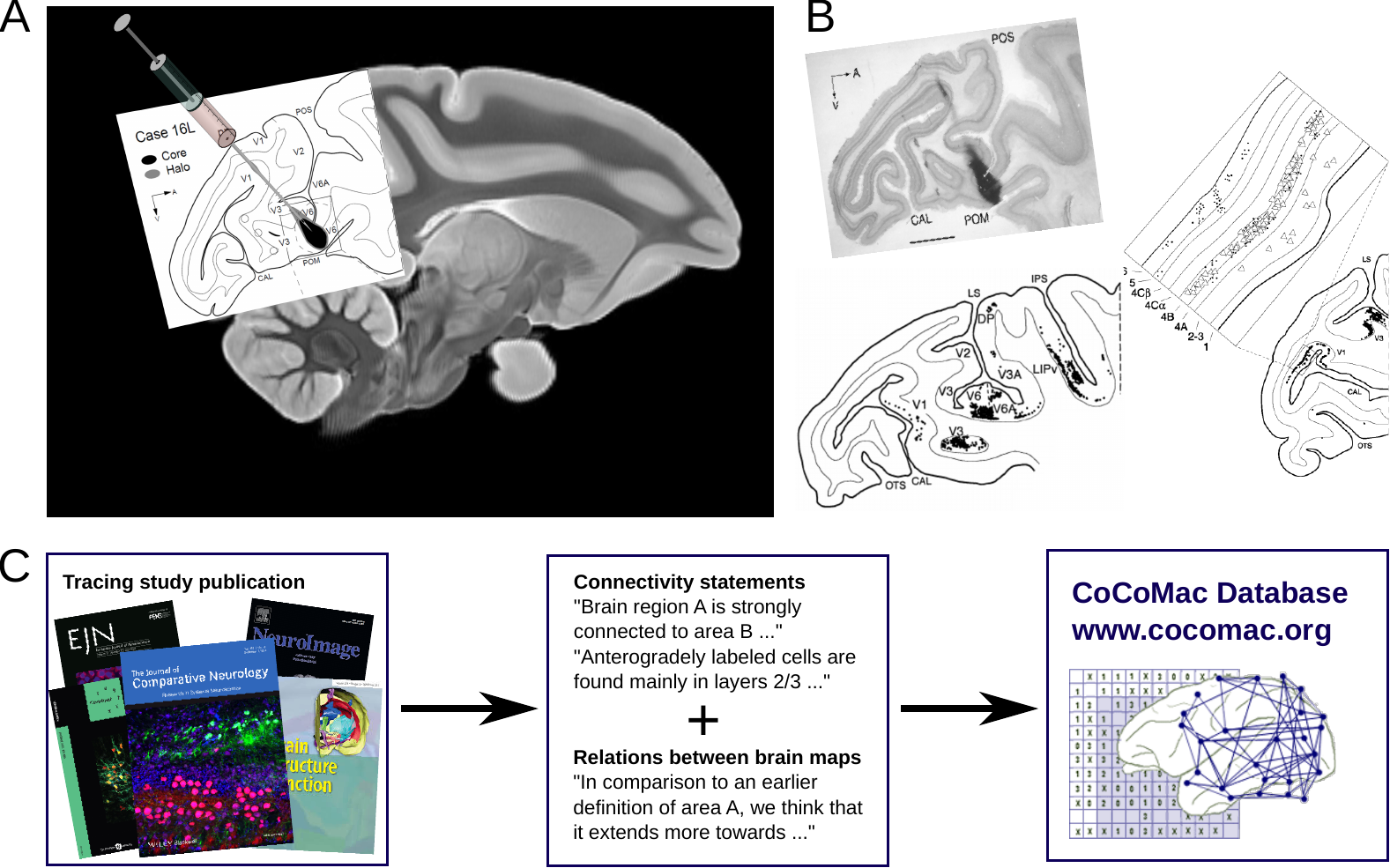}

\caption{Workflow from tracing experiment to entry in the CoCoMac database.
\textbf{A.} A tracing study is performed to study a particular part
of the brain, by injecting a tracer substance into a target area.
Shown is Case 16L from Galletti et al. \cite{Galletti2001_1572},
here registered to the macaque atlas of Calabrese et al. \cite{Calabrese15_408}
via the Scalable Brain Atlas \cite{Bakker15}. \textbf{B.} Tracer
is picked up by axons, and depending on the substance it is either
transported anterogradely towards the axon terminals, or retrogradely
to the cell bodies, or both. After sacrificing the animal, a careful
investigation of labeled cell bodies and/or axon terminals across
the brain is carried out, sometimes including layer-specific quantitative
data. \textbf{C.} After the results have been written up and subject
to peer review, collators from the CoCoMac database take out statements
on connectivity and the definitions of brain areas.}

\label{fig:cocomac}
\end{figure}
The CoCoMac database, which stands for Collation of Connectivity data
on the Macaque brain \cite{Stephan01_1159,Bakker12_30}, contains
both anterograde and retrograde tracing data from a large number of
published studies, especially for the cerebral cortex, in part with
laminar resolution. Figure \ref{fig:cocomac} illustrates the prerequisites
for creating such a database. Another collation effort \cite{Scannell1995}
has reconstructed the area-level structural connectome of the cat
from qualitative axonal tracing data. The Allen Institute provides
an anterograde tracing data set encompassing hundreds of injections
throughout the mouse brain \cite{Oh2014}. A comprehensive characterization
of laminar target patterns of connections between cortical areas in
primate is missing to date.

Axonal tracing is a reliable method for identifying actual connection
pathways, and often serves as the ground truth for evaluating diffusion
tensor imaging results (cf. \subsecref{Diffusion-tensor-imaging}).
However, the fact that connectomes based on tracing data are a composite
of connectivity in many individuals warrants special caution in their
interpretation. The average or union of the connections in many brains
in all likelihood does not accurately represent the connectivity of
any individual brain.

\subsection{Diffusion tensor imaging (DTI)\label{subsec:Diffusion-tensor-imaging}}

Diffusion tensor imaging (DTI) is a form of diffusion MRI or diffusion-weighted
imaging (DWI), which measures the local rate of water diffusion at
a resolution of typically a few millimeters.  DTI detects anisotropies
in the diffusion of water by using several different orientations
of the magnetic field gradients to obtain information about the directionality
of the diffusion in each voxel \cite{Basser94_259}. Since the diffusion
is greater along than perpendicular to myelinated axons, the method
enables the main local orientation of axonal fiber tracts to be identified.
The paths of the fiber tracts maximally consistent with the local
orientations are reconstructed using so-called tractography. The density
of these `streamlines' is a measure of connectivity between distant
brain regions, and can for instance be summed within cortical areas
to obtain an area-level cortical connectivity map. DTI is non-invasive
and can reveal the connectivity of the whole brain at once. However,
apart from possible directional specificity introduced by the choice
of seed points for tractography, the connectivity provided by DTI
is symmetric, as it can resolve the orientation but not the direction
of fiber tracts. While most cortical inter-area projections are reciprocal
with positively correlated connection density in the two directions
\cite{Markov2014_17,Beul15_3167,Majka20_1}, a proportion of connections
is asymmetric, and these asymmetries are hereby missed. Such asymmetries
are likely to be important for the dynamics predicted from neuronal
network models \cite{Knock09_86}. Further drawbacks of DTI are its
lack of laminar resolution and its inability to distinguish fibers
with different orientations in the same voxel, such as crossing or
touching (`kissing') fibers. Local tractographic errors due to kissing
or crossing fibers add up over distance, limiting the reliability
of the resulting connectivity maps, especially giving many false positives
for long-distance connections \cite{MaierHein17_1349}.

The Human Connectome Project provides high-resolution preprocessed
human diffusion MRI data for $>1100$ subjects. Tractography was performed
on an earlier, smaller data set from the Human Connectome Project
and the resulting connectome was made available via the Brainnetome
Atlas \cite{Fan16_3508}. Prominent DTI connectomes for the macaque
and mouse brains were published by Duke University \cite{Calabrese15_408,Calabrese15_bhv121}.

As yet, there is no straightforward way to derive fully reliable and
accurate connectomes from DTI. The same holds more generally for all
the types of connectivity information we have discussed. All experimental
connectivity data have `gaps': they only cover a certain spatial scale,
they represent a subsample or lack precision at the given scale, or
additional information is required to turn the experimental values
into numbers of synapses. For this reason, methods are needed for
filling in the gaps in the data in order to fully specify network
models. This is the topic of the next section.

\section{Predictive connectomics}

Where the experimental connectivity data have gaps, we can try to
fill these in using statistical estimates based on relationships of
the known connectivity with properties such as cytoarchitecture or
distance between brain regions. We refer to this approach as `predictive
connectomics'. Such statistical estimates still tend to have a high
degree of uncertainty associated with them, but if we want to fully
define a network model, there is no way around making certain assumptions
and approximations. From another perspective, the statements of predictive
connectomics represent formalized hypotheses for further anatomical
studies. The spatial and temporal organization of neurodevelopment
simultaneously explains many empirical relationships between connectivity
and other structural properties of the brain. In the present section,
we discuss the major heuristics for predicting connectivity, including
Peters' rule, architectural principles, and methods based on distance
and network topology, and describe how developmental origins form
a common denominator for many of these heuristics. Finally, we touch
upon the inference of structural connectivity from activity data.

\subsection{Peters' rule\label{subsec:Peters'-rule}}

\begin{figure}[h]
\includegraphics{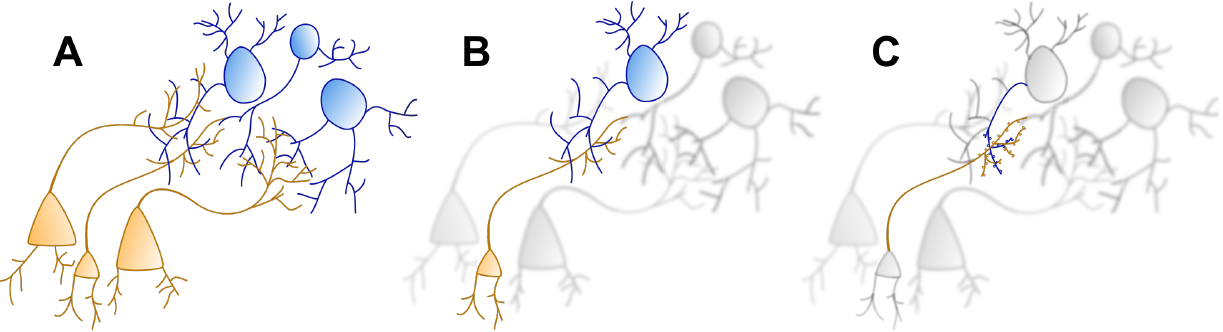}\caption{Illustration of the different levels of detail in the usage of Peters'
rule, as described in \cite{Rees17_63}. \textbf{A.} Population level,
\textbf{B.} Single-neuron level and \textbf{C.} Subcellular level.}
\label{fig:peters_rule}
\end{figure}

Peters' rule postulates that proximity between neurites (i.e. presynaptic
axons and postsynaptic dendrites) can predict neuronal connectivity.
It was originally proposed by Peters and Feldman (1976) \cite{Peters76_63}
for the projections from the lateral geniculate nucleus to the visual
cortex of the rat. The term `Peters' rule' was later coined by Braitenberg
and Schüz (1991) \cite{Braitenberg91}, who also generalized this
idea beyond the particular case studied by Peters and Feldman. The
rule has since been widely used by researchers. Over time its application
has varied. Rees et al. (2017) \cite{Rees17_63} reviewed the relevant
literature and distinguished between three conceptually different
usages of the rule, which correspond to increasing level of detail
(illustrated in \figref{peters_rule}):
\begin{enumerate}
\item Population level. In the original formulation, the rule was applied
as a predictor of connectivity between populations of neurons of the
same type. Consider a group of neurons A (for example in the thalamus)
projecting to a region containing another group B (for example pyramidal
cells in visual cortex), where all neurons within the groups are of
the same type. According to the original rule, the number of synapses
between A and B is correlated with the spatial overlap of presynaptic
axons of population A and postsynaptic dendrites of population B.
\item Single-neuron level. Extending the example from the previous point,
take two neurons $a_{i}$ and $b_{j}$ from populations A and B, respectively.
In this formulation, the probability $p_{ij}$ for a connection between
$a_{i}$ and $b_{j}$ to exist is proportional to the spatial proximity
between their respective pre- and postsynaptic arbors.
\item Subcellular level. At the subcellular level, Peters' rule has been
used to link the number of axonal-dendritic appositions to the number
of synapses, regardless of cell types.
\end{enumerate}

Peters' rule is not universal and has been shown to hold for certain
cases and fail in others, for all levels of detail. Section \ref{subsec:Microscopy}
describes an exception to Peters' rule at the subcellular level, which
probably carries over to the single-neuron level as well: an apposition
is more likely to predict a synapse if other synapses are present
on the same neurites \cite{Kasthuri2015_648}. Other studies have
provided evidence both in favor of and against the heuristic at the
subcellular level \cite{Packer13_2790,MerchanPerez2014_1579,Lee2016_370,Motta2019_eaay3134}.
Neurite proximity is undeniably a necessary condition for the formation
of synapses, but in general not sufficient to explain it, for instance
as activity-dependent plasticity may support preferential connectivity
between neurons with similar response properties. Nevertheless, Peters'
rule is a decent heuristic at the population level, with the main
caveat that some cell types do not connect to each other even if they
come into close proximity \cite{Binzegger04,Rees17_63}. Thus, the
rule may be fruitfully applied at the population level as long as
such cell-type-specific absence of connections is taken into account.

\subsection{Architectural principles}

The cytoarchitecture and laminar composition of cortical areas are
predictive of their connectivity, as first noted for frontal areas
of macaque cortex \cite{Barbas86_415,Barbas97}. In particular, architecturally
more similar areas are more likely to be connected, and if they are
connected, the connection density tends to be higher \cite{Hilgetag10_1006,Beul15_3167,Hilgetag16,Beul17}.
However, while architectural similarity reliably predicts the existence
and absence of connections, connection densities are better explained
by inter-area distances (cf. \subsecref{Distance-dependence}) \cite{Hilgetag10_1006}.
The characterization of areal architecture in terms of laminar differentiation
was systematized using the notion of architectural types, which also
consider the thickness of layer IV \cite{Dombrowski01}. Areas with
low architectural type have low neuron density, a thin or absent layer
IV, and indistinct lamination. Areas with high architectural type
have high neuron density, a thick layer IV, and distinct lamination.
The progression from low to high architectural types roughly corresponds
to the inverse of cortical hierarchies, down from limbic to early
sensory areas. Instead of using architectural types, which discretize
what is in fact a continuum of structural features across areas \cite{vonEconomo27},
one may use neuron density as a continuous explanatory variable. However,
compared to neural density differences, architectural type differences
are a better predictor of the existence and absence of connections
between macaque visual areas \cite{Hilgetag16}.

\begin{figure}
\includegraphics[width=1\textwidth]{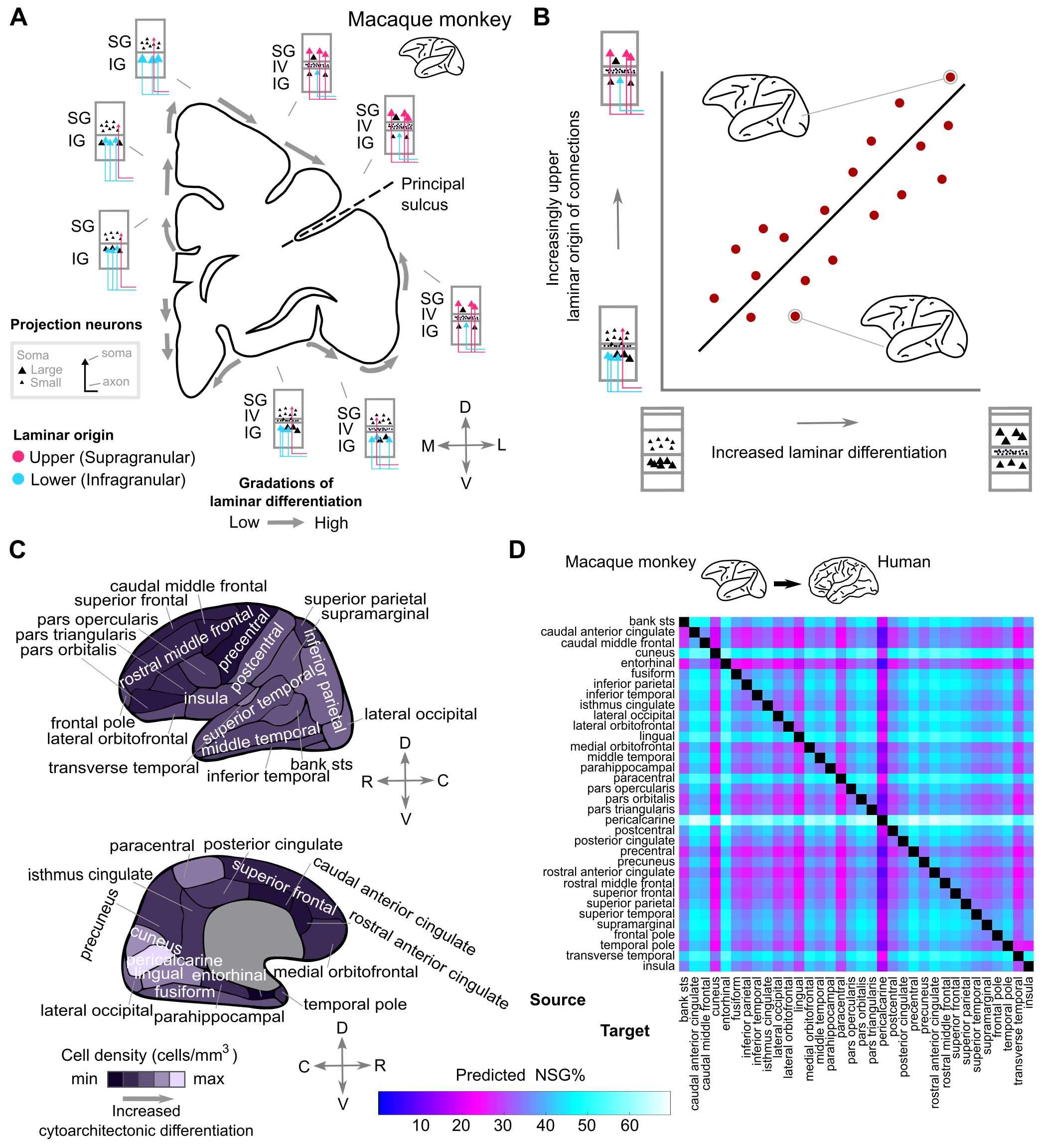}\caption{Laminar origin of connections, cytoarchitecture, and predictive connectomics.
\textbf{A.} Laminar origin of connections shifts from lower to upper
layers across the cortical sheet of the macaque monkey. \textbf{B.}
Schematic illustration of the quantitative relation between the cytoarchitecture
of cortical areas and the laminar origin of their connections to other
areas. The transition from less to more laminar differentiation (horizontal
axis), associated also with increased neural density, is accompanied
by a transition of predominantly lower to upper laminar origin of
connections (vertical axis). \textbf{C.} Cell densities of human cortical
areas based on von Economo and Koskinas, 1925 \cite{vonEconomo1925}.
Top, lateral view and bottom, medial view of the right hemisphere.
\textbf{D.} A monkey-to-human prediction of laminar origin of connections
($NSG\%,$ relative number of supragranular neurons) between all pairs
of cortical areas based on human cell densities (Panel C) and the
relationship between cytoarchitecture and the laminar origin of connections
(Panel B). Panel A based on a drawing from \cite{Sanides70_137}.
Panels C and D reproduced from \cite{Goulas2019_e2005346}.}
\label{fig:predictive_connectomics}
\end{figure}

Besides correlating with the existence or absence of connections and
with connection density, architectural differences are informative
of laminar projection patterns. Cytoarchitectonic difference is the
only consistent predictor that explains the majority of the variance
in laminar source patterns when compared with other candidate explanatory
variables such as rostrocaudal distance \cite{Goulas2019_e2005346}.
Areas with more distinctive layers and higher neuron density tend
to send projections from their upper (supragranular) layers to areas
with less distinctive layers and lower neuron density. Reversely,
projections from the latter to the former type of areas tend to emanate
from the lower (infragranular) layers. These patterns seem to generalize
across species, having already been demonstrated for cat, marmoset,
and macaque \cite{Goulas18_775}. Since laminar origin patterns are
correlated with laminar termination patterns, for instance supragranular
projections tend to target the granular layer IV \cite{Felleman91_1},
also termination patterns can be in part inferred from architectural
similarity \cite{Beul15_3167,Schmidt18_1409}. However, as the majority
of layer-resolved axonal tracing data is retrograde, origin patterns
have been more extensively studied than termination patterns. For
human cortex, laminar origin and termination patterns of inter-area
projections are still mostly unknown. For modeling purposes, the relationships
between laminar patterns and cytoarchitectural differences between
areas that have been observed in different mammalian species may be
used to assign laminar patterns to human connectomes (\figref{predictive_connectomics}).

Cortical thickness similarity has also been investigated as an explanatory
variable for inter-area connectivity. Areas with more similar thickness
are more likely to be connected, although this relationship does not
hold consistently \cite{Hilgetag16}. Thickness differences also relate
to laminar patterns: projections from thinner to thicker areas tend
to have a more supralaminar origin \cite{Beul17}. The fact that cortical
thickness is somewhat predictive of connectivity fits with the observation
that cortical thickness correlates negatively with neuron density
\cite{Beul17,Schmidt18_1409}. However, compared to cortical thicknesses,
architectural types and neuron densities are more systematically related
to connectional features, indicating that cytoarchitecture is at the
heart of the relation between cortical thickness and connectivity.
More commonly, thickness similarity has been characterized in the
sense of co-variation across subjects, areas with positively co-varying
thicknesses across subjects being more likely to be connected \cite{Lerch06_993,Gong12_1239,Alexander-Bloch13_322}.
However, also this correlation is far from perfect, and a large percentage
of regions have co-varying thickness without being connected \cite{Gong12_1239}.

\subsection{Distance dependence\label{subsec:Distance-dependence}}

Both for connectivity between neurons within a given brain region
and for that between brain regions, shorter connections are more likely
or more numerous than longer ones. This rule makes sense considering
the material and energetic cost of wiring and the space taken up by
axons and axon bundles. Nevertheless, non-random long-range connections
between specific regions exist, which are in part explained by spatiotemporal
patterns of brain development (cf. \subsecref{Neurodevelopmental-underpinnings}).
Locally within cortical areas, connection probability of both excitatory
and inhibitory neurons falls off approximately exponentially with
intersomatic distance with a space constant around $150-300\:\mu\mathrm{m}$
\cite{Song05_0507,Perin11,Packer2011_13260,Levy2012_5609}. Besides
these local connections, pyramidal cells establish patchy connectivity
at distances on the scale of millimeters \cite{Voges08}.

Similarly to local connectivity, projections between cortical areas
follow an `exponential distance rule' in which the lengths of axons
are exponentially distributed and the probability for a neuron to
send a projection between cortical areas thus falls off exponentially
with distance \cite{ErcseyRavasz13}. This exponential distance rule
at the level of individual neurons translates into an exponential
decay in connection density at the level of areas as well \cite{Schmidt18_1409}.
Given the connectivity between cortical areas, the spatial arrangement
of areas in the brain to a good approximation minimizes the total
wiring length \cite{Young92_152,Klyachko03_7937,ErcseyRavasz13}.
In a study of the connectivity between macaque cortical areas \cite{Markov2014_17},
the combination of the log ratio of neuron densities and Euclidian
distance between areas provided the best statistical predictions of
the existence of connections \cite{Beul17}. All in all, physical
distance constitutes a useful explanatory variable for the existence
and density of both local and long-range connectivity.

\subsection{Connectome topology\label{subsec:Connectome-topology}}

So far we have considered connectivity predictions based on the properties
of pairs of network nodes (neurons or areas). It is possible to go
beyond pairwise properties and look at patterns of three or more nodes
to infer connectivity. According to the homophily principle---described
in social network theory as `the tendency to choose as friends those
similar to oneself' \cite{Granovetter83_201}---nodes with common
neighbors are more likely to be themselves connected \cite{Goulas2019_eaav9694,Goulas2019_e2005346}.
This property is for instance displayed by so-called small-world networks,
in which a combination of many short-range and a few long-range connections
enables any node to be reached via a small number of hops through
the network. The homophily principle holds sway both at the single-neuron
level and at the level of brain regions, in both vertebrate and invertebrate
brains \cite{Goulas2019_eaav9694}.

In local cortical circuits, certain connection motifs---patterns
of connectivity in small groups of nodes---between three or more
neurons are overrepresented with respect to random graphs defined
by pairwise connection probabilities alone \cite{Song05_0507,Perin11}.
In a study of groups of up to twelve neurons, the probability of a
connection between a pair of neurons was found to increase linearly
with the number of common neighbors. Through this expression of the
homophily principle, cortical neurons cluster into small-world networks
\cite{Perin11}. Furthermore, like-to-like connectivity between neurons
with similar functional specificity, e.g., neurons in primary visual
cortex having similar orientation preference or responding to the
same type of visual stimuli \cite{Ko2011_87}, is an important ingredient
of the local network topology \cite{Billeh2020}.

\begin{figure}[h]
\includegraphics[width=1\textwidth]{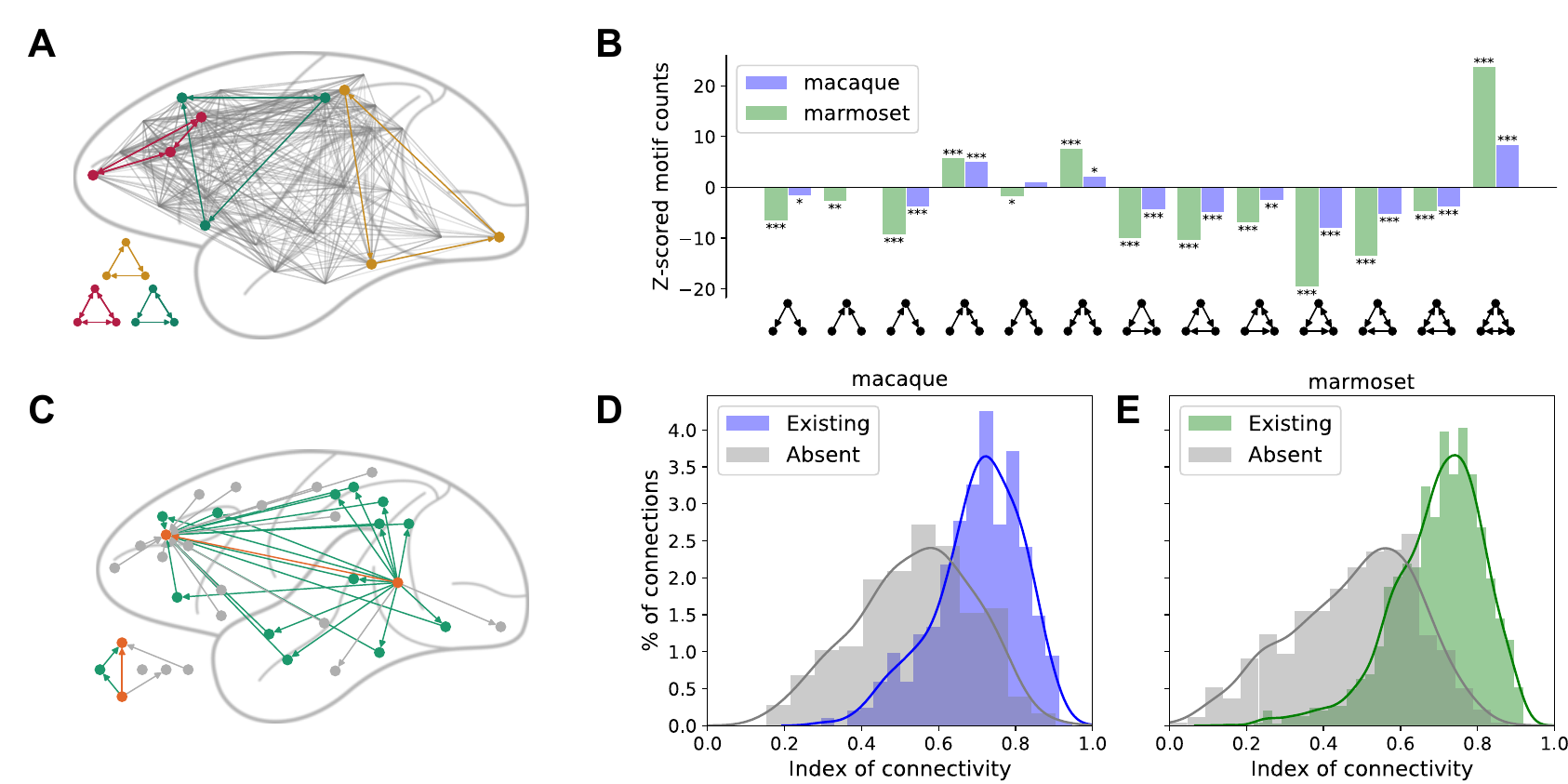}

\caption{Illustration of topological connectivity features of macaque and marmoset
cortical graphs. \textbf{A. }Schematic depiction of motifs in the
area-level macaque cortico-cortical connectivity.\textbf{ B. }Z-score
of the motif counts for all connected triads in the macaque and marmoset.
Motif counts are normalized by the mean and standard deviation of
the motif counts from $1,\!000$ random graphs with the same connection
probability as the experimental data in each case; {*} $p<0.05$,
{*}{*} $p<0.01$, {*}{*}{*} $p<0.001$. \textbf{C.} Schematic depiction
of the area-level index of connectivity as described in \cite{Jouve98_28}.
Shared neighbors (green nodes) contribute to the prediction of a direct
connection (orange), while non-shared neighbors (gray nodes) make
a direct connection less likely. \textbf{D,E.} Distribution of the
index of connectivity for existing and absent cortico-cortical connections
in macaque (D) and marmoset (E).}

\label{fig:Jouve}
\end{figure}

At the level of brain regions, Jouve et al. \cite{Jouve98_28} noticed
that directly connected areas in macaque vision-related cortex have
far more indirect connections between them than do unconnected areas.
The author defined an index of connectivity that captures the fraction
of shared first-order intermediate nodes between any two areas (Figure
\ref{fig:Jouve}A). They found that this metric is related to the
existence or absence of connections in macaque visual cortex, and
used this to infer the connectivity of area pairs for which no tracing
data were available. As pointed out in the study, the given indirect
connectivity index cannot predict all connections accurately, but
nevertheless exposes an underlying principle in the structure of the
primate connectome.

We computed the index of indirect connectivity and the triadic motif
counts on the tract-tracing data from macaque \cite{Markov2014_17,Mejias2016}
and marmoset \cite{Majka20_1} monkeys, using the subgraphs without
unknown connections. This analysis reveals that the motif counts,
relative to random graphs defined by pairwise connection probabilities
alone, have a similar structure in both primates, as shown previously
\cite{Theodoni2020} (Figure \ref{fig:Jouve}B). We also see that
the index of connectivity has a large overlap for areas with and without
a direct connection in both primates (Figure \ref{fig:Jouve}D). However,
extreme values ($>0.8$ and $<0.3$) reliably distinguish existing
connections from non-existing ones. 

A combination of spatial proximity and homophily accounts for many
topological characteristics of human cortical networks such as degree,
clustering, and betweenness centrality distributions \cite{Vertes2012_5868,Betzel2016_1054}.
Chen et al. (2020) \cite{Chen2020_bhaa060} found that adding cytoarchitectonic
similarity to distance dependence and topological constraints resulted
in even better predictions when applied to the macaque cortical connectome.
These findings place local topology, and especially homophilic attachment,
in the list of overarching properties governing neural network structure.

\subsection{Neurodevelopmental underpinnings of connectivity heuristics\label{subsec:Neurodevelopmental-underpinnings}}

Many of the aforementioned connectivity heuristics can be brought
together in a common developmental framework. The spatiotemporal ontogeny
of the brain provides simultaneous explanations for distance-dependent
connectivity, the preferential connectivity between cytoarchitectonically
similar areas, and aspects of the network topology of the brain \cite{Goulas2019_eaav9694}.
It also accounts for deviations from a simple decay in connection
probability with distance. For instance, changes in the parameters
of the distance-dependent connectivity during development can yield
a small-world network structure with multiple clusters \cite{Nisbach07_185}.
Limbic cortical areas, of low architectural type, develop earlier
and over a shorter period than areas of high laminar differentiation
such as primary visual cortex. This rapid development not only underlies
the less distinct lamination and low neuron density of limbic areas,
but also gives these areas a longer time window for connecting to
other regions, thus supporting their coordinating role \cite{Barbas16_125}.
The importance of spatial embedding and heterochronicity---the existence
of a sequence of developmental time windows---for brain wiring were
demonstrated for species ranging from the fruit fly to the mouse,
rat, macaque monkey, and human \cite{Bayer87_57,Goulas2019_eaav9694}.
Thus, taking into account spatiotemporal gradients of brain development
can help predict more realistic connectomes regardless of the species
under investigation.

\subsection{Reconstructing connectivity from activity}

So far we have focused on predictive relations derived from the anatomical
features of the nervous tissue. However, anatomical information is
often costly to obtain or requires invasive methods and is therefore
often not available for all the different brain regions. An alternative
approach is to derive neural network structure from activity data.
While promising results in this direction have been obtained, this
approach suffers from the drawbacks that widely different network
parameters can lead to closely similar activity \cite{Prinz-2004_1345}
and that the external input to the network modulates the link between
structure and activity \cite{Aertsen89}.

When relating activity to connectivity, we need to distinguish a few
different terms. Besides structural connectivity, the topic of this
chapter, there are two types of activity-dependent `connectivity':
so-called functional connectivity, and effective connectivity. Functional
connectivity is symmetric between source and target nodes, and describes
correlations between their activity. It is often used in the context
of functional imaging studies to characterize the interactions between
brain regions. Effective connectivity is a directed measure, describing
the minimal graph that would be needed to account for the observed
interactions between nodes \cite{Aertsen89}. In a stricter mathematical
sense, one can define effective connectivity as the product of the
structural connectivity and effective synaptic weights that depend
on the activity level of the target nodes and quantify their susceptibility
to increased input \cite{Albada15}. The same structural substrate
can support different functional and effective connectivities depending
on the external drive and the network state. When inferring structural
connectivity from activity data, the lines between the different types
of connectivity can be somewhat blurred, but it is useful to keep
in mind the distinctions.

We have already discussed two physiological methods that help estimate
structural connectivity at the microscopic scale: paired recordings
(\subsecref{Paired-recordings}) and glutamate uncaging (\subsecref{Glutamate-uncaging}).
These methods provide reliable connectivity data, but are constrained
to small numbers of neurons. Parallel electrophysiological recordings
of up to hundreds of individual neurons are now possible for instance
with Utah arrays or Neuropixels probes \cite{Maynard97,Jun2017},
and functional magnetic resonance imaging enables recording whole-brain
activity, resolved into ever smaller voxels \cite{Zimmermann2011_e28716,DeMartino2013_e60514}.

A number of methods have been proposed for inferring the underlying
connectivity from these large-scale activity data. Time-lagged correlations
between the spike trains of pairs of neurons are informative about
the direction of the information flow and have been shown to be linked
to the structural connectivity \cite{Ostojic09_10234}. A few studies
have used this fact to reconstruct network connectivity from parallel
spike train cross-correlation histograms \cite{English2017,Pastore2018,Kobayashi19_1}.
Pairwise correlations are shaped not only by direct connections between
neurons, but also by indirect connections, the electrophysiological
properties of the individual neurons, transmission delays, and the
external drive to the network \cite{Cohen11_811,Helias13_023002,Helias14}.
Given certain conditions such as stationarity and knowledge of the
single-neuron electrophysiology, the structural connectivity can in
principle be uniquely reconstructed from the pairwise correlation
functions; that is, one can compute and thereby take into account
the influence of the indirect connections and shared input \cite{Grytskyy13_NWG,Helias14,Albada15}.
In practice, biological neural networks do not fulfill ideal conditions
and experiments do not fully provide the required information, setting
a ceiling on the accuracy of structural connectivity inferred from
correlations.

Going beyond pairwise correlations, Casadiego et al. \cite{Casadiego2018}
propose a method for inferring synaptic connections from the dependence
of inter-spike intervals on cross-spike intervals, i.e. intervals
between the spike times of different neurons. The method can successfully
distinguish excitatory and inhibitory synapses, as validated with
point neuron network simulations. Networks exhibiting phase-locked
activity may not sufficiently explore the dynamical landscape to enable
all synapses to be reconstructed. In such cases it can help to expose
the network to different external driving conditions \cite{vanBussel11}.
Similarly using only knowledge of the spiking activity and not requiring
membrane potential traces, Zaytsev et al. \cite{Zaytsev2015} infer
the connectivity of simulated networks of a thousand neurons using
maximum likelihood estimation of a generalized linear model of the
spiking activity. Such methods based on generalized linear models
can work well when the activity of all neurons is recorded \cite{Gerhard2013_e1003138},
but, like for any connectivity reconstruction method, undersampling
is expected to diminish their performance.

Fitting the observed activity to a dynamical network model can be
a complex and computationally intensive procedure. Structural connectivity
parameters are sought that optimize a score or cost function based
on some features of interest. In simulation-based methods, optimal
parameter combinations can be searched via brute force \cite{Prinz-2004_1345,Stringer2016},
stochastic optimization techniques such as evolutionary methods \cite{Druckmann07_7,Rossant10,Carlson2014},
or plasticity rules \cite{Diaz16_57}. Likelihood-based methods do
not require costly simulations \cite{Paninski04,Pillow2005,Ladenbauer2019,Rene2020_1448}
and under some conditions allow straightforward optimization via gradient
ascent or simplex methods. However, estimating the analytical likelihood
function is a challenging task for complex models. Machine learning
methods are starting to be developed that can overcome this issue
and estimate parameter distributions given emergent dynamical properties
of modeled networks \cite{Bittner2019,Goncalves2020}.

All in all, establishing unequivocal links between structural connectivity
and neural activity remains a major challenge in neuroscience, and
structural connectivity estimates from population recordings should
generally be interpreted with caution.

\section{Validation of predicted connectivity}

The most direct way of validating connectivity predictions is of course
experimental confirmation. Barring the ideal situation where this
is possible, we have a few options at our disposal for putting predictions
to the test. In this context, different types of predictions exist:
sometimes, a full connectome is generated, while sometimes merely
statistical regularities in connectivity data are obtained. For the
case of full connectomes, we can further distinguish generative models
that do not directly rely on connectivity data, for instance based
on distance, cytoarchitectonics, and topological constraints; and
cases where gaps in connectivity data are filled in.

Where the result of the prediction is a full connectome, one can compare
with experimentally obtained connectomes either edge-wise or based
on graph properties such as degree distributions, clustering, modularity,
characteristic path length, small worldness, or betweenness centrality
\cite{Vertes2012_5868,Betzel2016_1054}. The choice of properties
to compare is nontrivial and depends on their presumed importance
with regard to the scientific question. Ideally, the fitness of the
generative model is quantified using a likelihood function, but where
this is difficult, other objective functions may be defined \cite{Betzel2017_20170623}.

In case of statistical fits to connectivity data, we can check the
robustness of the predictions by determining confidence intervals
for the fit parameters. When no straightforward expressions for these
are available, bootstrapping provides a solution in which random data
samples are drawn with replacement and the statistic of interest is
computed for each sample \cite{Mooney1993}. A similar strategy can
be applied when filling gaps in connectomes: leaving out part of the
known data and either determining how well the predictions fit to
the left-out data, or again computing graph properties and assessing
their variability. Alternatively, we can add noise to the underlying
data on the order of the uncertainty in the data. Depending on the
case, `uncertainty' in this context can for instance include experimental
noise, inter-individual and inter-species variability, or uncertainty
due to mapping between parcellations. Since it is in practice difficult
to determine the size of the uncertainty, one can add different levels
of noise to the estimated model parameters and check whether the predictions
hold true even for relatively high noise levels.

Another route for testing the plausibility of connectivity predictions
is to build corresponding network models, perform dynamical simulations,
and compare the resulting activity with experimental activity data.
Software tools supporting the systematic comparison between simulated
and experimental activity data are available for both single neurons
and networks of neurons \cite{Gutzen18_90}. This method is complicated
by the fact that not only the connectivity but also the dynamical
properties of the nodes (neurons or populations of neurons), the transmission
delays, and the external drive contribute to the network dynamics.
However, depending on the dynamical regime, network dynamics can be
fairly robust to electrophysiological properties of the individual
nodes \cite{Sahasranamam2016}. The parameter space can be explored
systematically via parameter scans, or in a more targeted manner via
stochastic optimization. If at least some parameter settings for
the nodes, delays, and external drive, consistent with biological
data, can be found for which the predicted connectivity yields realistic
activity, this provides some degree of validation. Stronger support
is provided if the experimental activity data are no longer successfully
reproduced upon changing the connectivity. Ultimately, neural network
models should be consistent with both anatomical and electrophysiological
properties of the brain.

\section{Concluding remarks}

Data on brain anatomy are increasingly made available as systematic,
quantitative data sets, facilitating their use in neuronal network
models. Inspired by seminal works like those of von Economo \cite{vonEconomo1925}
and Braitenberg and Schüz \cite{Braitenberg91}, modern anatomists
recognize the importance of systematization and quantification for
informing analyses and models. Historically, much anatomical data
was made available only in the natural language text of publications.
On the example of tracing studies, the creators of the CoCoMac database
\cite{Stephan01_1159} recognized the need to bring these data into
a machine readable format and to create a framework for systematically
mapping the parcellations mentioned in the text to different parcellations
of choice when constructing connectivity maps. The modern, systematic
way of publishing data is most prominently represented by large-scale
initiatives like the Allen Institute for Brain Science, Janelia Research
Campus, the Human Connectome Project, the Japanese Brain/MINDS project,
and the European Human Brain Project. Nevertheless, there is sometimes
still a disconnect between experimentalists and computational neuroscientists
in terms of the formats in which the data are published. Anatomical
data are still often made available as image files which require additional
processing before they can flow into models, in formats specific to
the discipline. An illustrative anecdote is that in 2018 Schmidt et
al. \cite{Schmidt18_1409} still obtained cortical thickness from
micrographs by measuring with a ruler the distance between layer markers.
One reason why modelers generally cannot use image data directly is
that they tend to work with concepts like definite cortical areas
and layers, rather than in a spatial continuum. These categorical
concepts constitute strong hypotheses that help to reduce and interpret
the data. Tables of area or laminar averages are then more useful
than images. If the data are offered as images, at least scripts and
documentation should be published alongside the data to enable the
relevant quantities to be potentially more easily extracted. The latter
approach retains flexibility with respect to particular parcellations
and is future-proof as algorithms of feature extraction improve and
concepts of brain organization may change over time.

We have described methods ranging from microscopy to diffusion magnetic
resonance imaging for measuring connectivity. However, this list is
not exhaustive and novel techniques are continuously developed. A
modern technique is polarized light imaging (PLI), which measures
fiber orientations in brain slices using the birefringence properties
of myelin \cite{Larsen07_851,Axer11}. Three-dimensional reconstructions
enable fiber tracts to be followed through the brain at a resolution
of some tens of micrometers. Axons entering the white matter can be
visualized with an in-plane pixel size down to the micrometer scale.
An add-on to PLI, also based on transmitting polarized light through
histological sections, is Diattenuation Imaging, which provides complementary
information on tissue composition \cite{Menzel19_1}. These methods
promise new ways of determining the connectivity of neural network
models.

Also in the field of predictive connectomics, our treatment of methods
has not been exhaustive. Besides predictions based on the proximity
of neural processes or cell bodies, cytoarchitecture, topological
constraints, and neural network activity, it is for instance possible
to generate connectomes based on gene expression data \cite{Fornito2019_34,Barabasi2020_435,Timonidis2020_1}.
Another possibility we have only briefly alluded to is a normative
approach, in which the connectome is in some sense assumed to be optimal,
and the implications of this assumption for connectivity are investigated
\cite{Chklovskii04_609,Samu2014_e1003557}. As in so many fields of
science, machine learning methods and artifical neural network models
provide a promising new avenue for identifying regularities in data
that help to predict connectivity.

As we have seen, connectomes for neural network models are subject
to a variety of uncertainties. Each experimental method carries with
it measurement errors, data from multiple individuals tend to be needed
to fully specify a connectome, and in many cases the best available
estimates even come from different species. We have largely skipped
over the vast and difficult topic of mapping data between species.
In many cases, the sobering truth is that this cannot be done in a
fully principled manner. All types of uncertainties, whether due to
experimental methods, individual differences, or interspecies differences,
lead to uncertainties in predicted model dynamics. We have described
some ways of verifying the robustness of network models to these uncertainties.

Brain models based on these statistical rules are necessarily models
of an average brain. This limits their explanatory power. Not only
in humans but also in other species, macroscopic features of brain
dynamics, like dominant frequencies and functional connectivity, vary
from individual to individual \cite{vanAlbada2007_279,Gordon2017_386,Xu2019_543}.
When the deviation of simulated brain activity from experimental data
is of the same order as the inter-individual variability, there is
nothing left to explain for this type of model. Schmidt et al. \cite{Schmidt18_e1006359}
illustrate this situation for the prediction of functional connectivity
between areas on the basis of a spiking network model. Such observations
challenge the research strategy to aggregate data from different species
and individuals to arrive at a statistical model of brain structure.
Progress may eventually only be possible by further constraining generic
connectivity rules by anatomical data obtained from the individual
delivering the brain activity data to be predicted \cite{Proix17_641}.

Ultimately, the statistical descriptions we apply to summarize brain
organization are not the rules by which brains are built in nature.
The rules mathematically formalize the limits of our knowledge on
the structure of individual brains. And using these rules is to date
just the most efficient way of instantiating large-scale neuronal
networks in a computer by a fully parallel process \cite{Morrison05a}.
In nature brains are pre-shaped by evolution and further formed by
growth rules in continuous interaction with the environment. Eventually
we need to understand and formalize these more fundamental rules to
grow artificial individual brains in a computer. This implies the
existence of a sufficiently accurate model of the environment. Averages
over such model instances then in turn need to be consistent with
our former statistical descriptions of brain structure.

Nevertheless, the major short-term challenge consists in the construction
of brain models encompassing different brain components, as already
alluded to in the introduction of this chapter. With a few notable
exceptions, until today models of neuronal networks are usually constructed
by a single researcher, often a PhD student, or small research groups.
It seems likely that we have hit a complexity barrier and for this
reason the complexity of the majority of models has not increased
much over past decade. In order to integrate the heterogeneity of
different brain areas and their multi-level hierarchical organization
into a brain model will require that we learn to use models of brain
components created by other researchers as building blocks.

International large-scale projects like EBRAINS have started to create
the ICT infrastructure enabling the sharing and reuse of data and
model components, as well as the simulation of multi-scale models
and their environments. The hope is that using these infrastructures
fosters the required culture of sharing and collaboration in neuroscience.

\section*{Acknowledgments}

Supported by the European Union\textquoteright s Horizon 2020 Framework
Programme for Research and Innovation under Specific Grant Agreements
No. 785907 and 945539 (Human Brain Project SGA2, SGA3) and Priority
Program 2041 (SPP 2041) \textquotedbl Computational Connectomics\textquotedbl{}
of the German Research Foundation (DFG).

\bibliographystyle{llcns/splncs}

\end{document}